# A Structured Approach to the Development of Solutions in Excel

Peter Bartholomew
MDAO Technologies Ltd
peter.bartholomew1@btconnect.com

**ABSTRACT**

*Spreadsheets offer a supremely successful democratisation platform, placing the manipulation and presentation of numbers within the grasp of users that have little or no mathematical expertise or IT experience. What appears to be almost completely lacking within a 'normal' solution built using Excel default settings is the deployment of any structure that extends beyond a single-cell formula. The structural elements that allow conventional code to scale without escalating errors appear to be absent. This paper considers the use of controversial or lesser-used techniques to create a coherent solution strategy in which the problem is solved by a sequence of formulas resembling the steps of a programmed language.*

## 1 INTRODUCTION

Spreadsheet technology, here the use of Excel is considered in particular, has achieved remarkable success in making 'numbers' available to business and industry with little or no reliance upon the user having any interest or knowledge in the mathematical theories or IT structures that make analysis possible in other domains. Indeed Raffensperger [2001, 2003] almost celebrates this lack of competence and concentrates upon cell-by-cell construction as a desirable characteristic, unlike Hellman [2005] who argued that the cell/ matrix formula concept is so deeply flawed that a separation of conceptual model from the presentation structure of the spreadsheet is long overdue. Raffensperger rejects the "metaphor" of the spreadsheet as a program and draws upon text documents, mathematics and graphic art for style guidance. In doing so, he chooses to ignore why modularity is so highly prized for code development or why the more abstract constructions of linear algebra are important in mathematics or even why technical reports tend to be structured using a hierarchy of numbered sections. Excel is superbly successful for the solution of *ad-hoc* problems but the structural elements that allow conventional code to scale without escalating errors appears to be sadly lacking in the spreadsheet solutions I have seen or even from the good-practice advice available within the Excel community.

The approach outlined in this paper started as an experiment conducted in response to a statement deprecating the use of Names in Excel and claiming their use to be limited to the "simplest" of applications[1]. The objective of the work is to examine whether that claim is true or whether it is possible to adopt the converse strategy and eliminate the use of direct cell referencing in its entirety. It turned out the use of Names for all referencing leads to the adoption of other techniques which, although standard features of Excel, tend not to receive extensive mainstream use. Overall the approach is more consistent with the programmatic approach of Bewig [2005] than it is to that of Raffensperger or the FAST Standard [2015].

The techniques adopted include the widespread use of array formulas, with the formula being decomposed into single-cell formulas through the use of relative referencing only where the use of array formulas proves impractical. Another feature of the approach is to use Named formulas (these do not refer to ranges) to reduce or eliminate the need to work with deeply nested formulas. This is shown to generate linear sequences of statements that, when documented, are far more reminiscent of a procedural language than of a typical worksheet formula but at the same time the immediacy of the spreadsheet solution and the opportunity for graphic presentation and user interaction are retained. It is only when one starts digging to see the formulas that the difference even becomes apparent; something that may affect the auditor but should not be obvious to the user.

[1] Names are better (or only) suited to simple spreadsheets with limited complexity, where reading a simple natural language formula such as *"= Price * Quantity"* is a real possibility

The change of mind-set embodied in the new approach also leads to changes within the formatting and presentation in order to reduce the emphasis on the location of data referenced within the worksheet and, instead, to stress the significance of the data as properties or components of a container data object. Because of this, there is little or no benefit in showing the normal alphabetic form for sheet column headers. The location of an object within the worksheet is simply not relevant to the solution so its position on the worksheet may be determined solely by aesthetic considerations and the need to achieve a clear visual presentation of content to the user.

Another feature of the approach is that, provided formulas are only entered as the properties (ideally the array property) of Named Ranges, the documentation of Names provides a complete description of the workbook. By 'complete' it is meant here that it is possible to rebuild the workbook in its entirety by using the documentation as steering data to direct VBA utility code. Only the input data and additional features such as charts or shapes and *ad-hoc* annotation would need to be provided by the developer.

## 2 DISCUSSION OF THE TECHNIQUES

### 2.1 Applicability

The approaches advocated and described in this paper relate to the building of spreadsheet-based numerical models and are far less applicable to the analysis of corporate data. In a model, the data may well have been created with the specific goal of animating the model and the data lifecycle tend to be one of archiving one set of data within a version of the model and then loading a fresh dataset into the model, although this might require some degree of modification re-sizing arrays to accommodate it.

This is a world away from the massive corporate databases that accumulate over time. Such data may accumulate in use but otherwise needs protection against unauthorised change if it is to retain its integrity. Eventually any such data-processing activity will crumble under the weight of accumulated data but the point may be deferred by rigorous optimisation for speed.

The author has experience of building models for simulating ship motion in the North Atlantic, calculating the strength and stiffness of carbon composite laminates, simulating the output from statistical analysis, animating the user interface for games such as 2048 and, more recently, financial modelling within a comparative study. The common feature of these models is that they are light on data but require multiple processing steps.

### 2.2 Example: Model with similar line items

The first example was a shared exercise conducted under Levi Bailey's leadership within a LinkedIn discussion. The solution shown in Figure 1 drew heavily upon techniques developed for problems far removed from financial modelling and was intended to provide an option within a spectrum of solutions, the principal ones being drawn from the finance modelling community. This solution has a number of highly distinctive features, which will be further discussed as topics within this section.

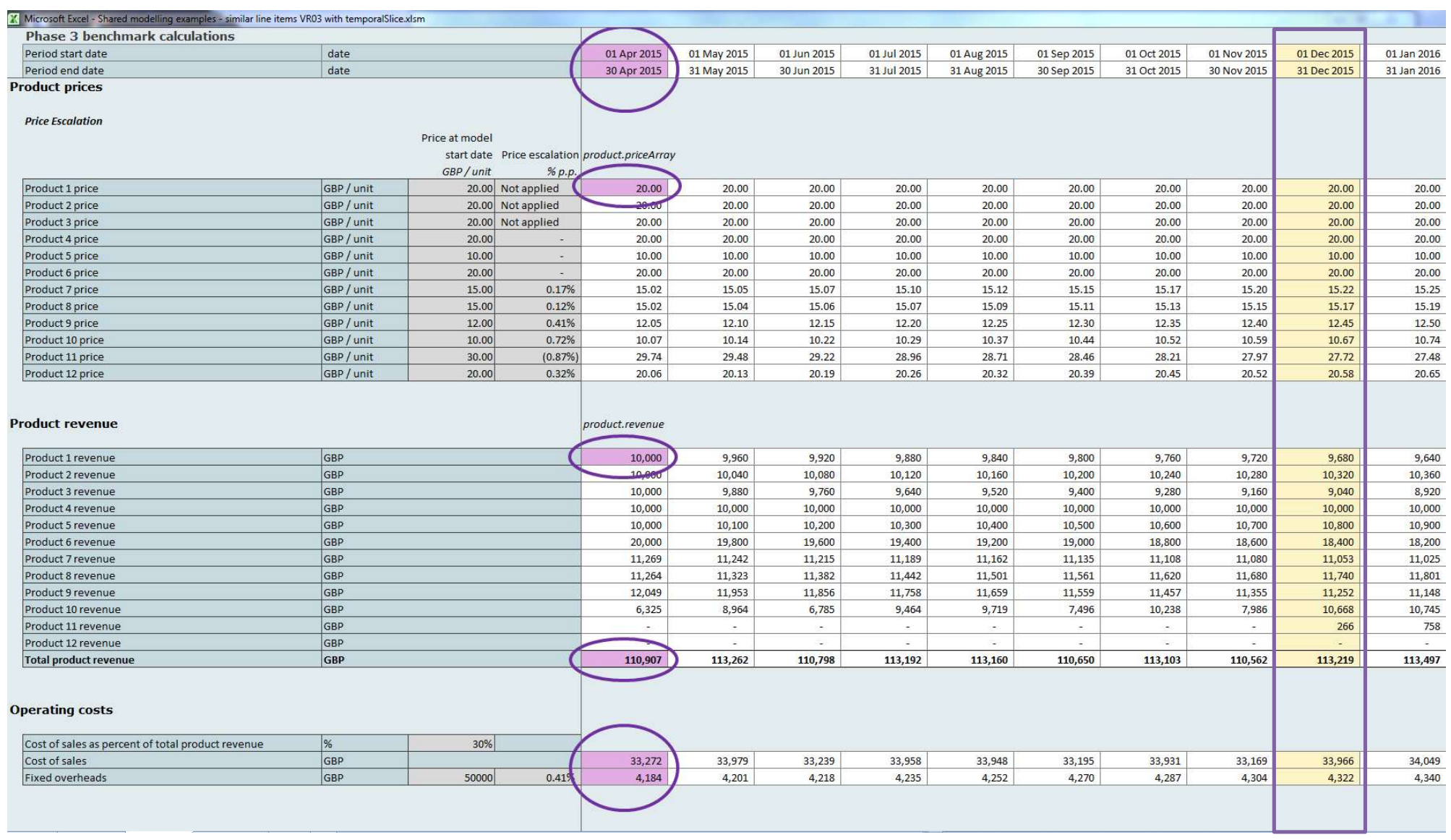

Microsoft Excel - Shared modelling examples - similar line items VR03 with temporalSlice.xlsm

Phase 3 benchmark calculations

| | | | | | | | | | | | | | |
|---|---|---|---|---|---|---|---|---|---|---|---|---|---|
| Period start date | date | | | 01 Apr 2015 | 01 May 2015 | 01 Jun 2015 | 01 Jul 2015 | 01 Aug 2015 | 01 Sep 2015 | 01 Oct 2015 | 01 Nov 2015 | 01 Dec 2015 | 01 Jan 2016 |
| Period end date | date | | | 30 Apr 2015 | 31 May 2015 | 30 Jun 2015 | 31 Jul 2015 | 31 Aug 2015 | 30 Sep 2015 | 31 Oct 2015 | 30 Nov 2015 | 31 Dec 2015 | 31 Jan 2016 |
| **Product prices** | | | | | | | | | | | | | |
| *Price Escalation* | | | | | | | | | | | | | |
| | | Price at model start date *GBP / unit* | Price escalation *% p.p.* | *product.priceArray* | | | | | | | | | |
| Product 1 price | GBP / unit | 20.00 | Not applied | 20.00 | 20.00 | 20.00 | 20.00 | 20.00 | 20.00 | 20.00 | 20.00 | 20.00 | 20.00 |
| Product 2 price | GBP / unit | 20.00 | Not applied | 20.00 | 20.00 | 20.00 | 20.00 | 20.00 | 20.00 | 20.00 | 20.00 | 20.00 | 20.00 |
| Product 3 price | GBP / unit | 20.00 | Not applied | 20.00 | 20.00 | 20.00 | 20.00 | 20.00 | 20.00 | 20.00 | 20.00 | 20.00 | 20.00 |
| Product 4 price | GBP / unit | 20.00 | - | 20.00 | 20.00 | 20.00 | 20.00 | 20.00 | 20.00 | 20.00 | 20.00 | 20.00 | 20.00 |
| Product 5 price | GBP / unit | 10.00 | - | 10.00 | 10.00 | 10.00 | 10.00 | 10.00 | 10.00 | 10.00 | 10.00 | 10.00 | 10.00 |
| Product 6 price | GBP / unit | 20.00 | - | 20.00 | 20.00 | 20.00 | 20.00 | 20.00 | 20.00 | 20.00 | 20.00 | 20.00 | 20.00 |
| Product 7 price | GBP / unit | 15.00 | 0.17% | 15.02 | 15.05 | 15.07 | 15.10 | 15.12 | 15.15 | 15.17 | 15.20 | 15.22 | 15.25 |
| Product 8 price | GBP / unit | 15.00 | 0.12% | 15.02 | 15.04 | 15.06 | 15.07 | 15.09 | 15.11 | 15.13 | 15.15 | 15.17 | 15.19 |
| Product 9 price | GBP / unit | 12.00 | 0.41% | 12.05 | 12.10 | 12.15 | 12.20 | 12.25 | 12.30 | 12.35 | 12.40 | 12.45 | 12.50 |
| Product 10 price | GBP / unit | 10.00 | 0.72% | 10.07 | 10.14 | 10.22 | 10.29 | 10.37 | 10.44 | 10.52 | 10.59 | 10.67 | 10.74 |
| Product 11 price | GBP / unit | 30.00 | (0.87%) | 29.74 | 29.48 | 29.22 | 28.96 | 28.71 | 28.46 | 28.21 | 27.97 | 27.72 | 27.48 |
| Product 12 price | GBP / unit | 20.00 | 0.32% | 20.06 | 20.13 | 20.19 | 20.26 | 20.32 | 20.39 | 20.45 | 20.52 | 20.58 | 20.65 |
| **Product revenue** | | | | *product.revenue* | | | | | | | | | |
| Product 1 revenue | GBP | | | 10,000 | 9,960 | 9,920 | 9,880 | 9,840 | 9,800 | 9,760 | 9,720 | 9,680 | 9,640 |
| Product 2 revenue | GBP | | | 10,000 | 10,040 | 10,080 | 10,120 | 10,160 | 10,200 | 10,240 | 10,280 | 10,320 | 10,360 |
| Product 3 revenue | GBP | | | 10,000 | 9,880 | 9,760 | 9,640 | 9,520 | 9,400 | 9,280 | 9,160 | 9,040 | 8,920 |
| Product 4 revenue | GBP | | | 10,000 | 10,000 | 10,000 | 10,000 | 10,000 | 10,000 | 10,000 | 10,000 | 10,000 | 10,000 |
| Product 5 revenue | GBP | | | 10,000 | 10,100 | 10,200 | 10,300 | 10,400 | 10,500 | 10,600 | 10,700 | 10,800 | 10,900 |
| Product 6 revenue | GBP | | | 20,000 | 19,800 | 19,600 | 19,400 | 19,200 | 19,000 | 18,800 | 18,600 | 18,400 | 18,200 |
| Product 7 revenue | GBP | | | 11,269 | 11,242 | 11,215 | 11,189 | 11,162 | 11,135 | 11,108 | 11,080 | 11,053 | 11,025 |
| Product 8 revenue | GBP | | | 11,264 | 11,323 | 11,382 | 11,442 | 11,501 | 11,561 | 11,620 | 11,680 | 11,740 | 11,801 |
| Product 9 revenue | GBP | | | 12,049 | 11,953 | 11,856 | 11,758 | 11,659 | 11,559 | 11,457 | 11,355 | 11,252 | 11,148 |
| Product 10 revenue | GBP | | | 6,325 | 8,964 | 6,785 | 9,464 | 9,719 | 7,496 | 10,238 | 7,986 | 10,668 | 10,745 |
| Product 11 revenue | GBP | | | - | - | - | - | - | - | - | - | 266 | 758 |
| Product 12 revenue | GBP | | | - | - | - | - | - | - | - | - | - | - |
| **Total product revenue** | **GBP** | | | **110,907** | **113,262** | **110,798** | **113,192** | **113,160** | **110,650** | **113,103** | **110,562** | **113,219** | **113,497** |
| **Operating costs** | | | | | | | | | | | | | |
| Cost of sales as percent of total product revenue | % | 30% | | | | | | | | | | | |
| Cost of sales | GBP | | | 33,272 | 33,979 | 33,239 | 33,958 | 33,948 | 33,195 | 33,931 | 33,169 | 33,966 | 34,049 |
| Fixed overheads | GBP | 50000 | 0.41% | 4,184 | 4,201 | 4,218 | 4,235 | 4,252 | 4,270 | 4,287 | 4,304 | 4,322 | 4,340 |

**Figure 1 Extract from “Shared Modelling Example with Similar Line Items”**

An immediate visual distinction is that the worksheet is presented without gridlines or sheet headings since the location or even the relative alignment of data is considered immaterial both to the business problem and to its solution. In the main, the visual appearance is produced by applying modified versions of the Microsoft built-in styles. An exception is the presence of somewhat lurid magenta cells (here circled), each one representing an array formula.

The region over which these identifiable array formulas are applied typically corresponds to a Named Range. The uniformity of a formula throughout the time series is generally considered good-practice but here it is actually enforced by the use of array formula and, as a result, consistency checks are not required; also there are very few formulas to verify as being correct. Moreover, a simple looking formula such as “*= Price * Quantity*” can be applied to calculate results across large regions within a workbook.

The other conspicuous visual feature is the vertical band of highlighted cells. This is created by conditional formatting used to show an entire column range that creates a temporal slice through the model allowing any aggregation of data that is used to characterise the entire model to be restricted to the selected time periods. It would be possible to achieve the same result using a system of flags and SUMIFS but that would lack the elegance of the range intersection solution that exploits the overall structure of the problem rather than developing multi-step formula solutions ‘bottom-up’.

## 2.3 Named Ranges

Whereas a programming language will use variables to allow the developer to reference values held in computer memory, a spreadsheet uses the location of cells presented within a rectilinear grid. In each case this represents a level of indirection; the developer has no need to know the actual location at which the value is held in the computer memory.

In the case of the code variable, the name chosen in any but the most primitive machine code is likely to provide some description of the intended content. Spreadsheets are different; the cell reference such as A1, simply aims to identify the content by its location within an apparent grid. Any meaning is normally to be inferred from annotation held in adjacent cells.

The use of defined names to reference the content comes as close to defining a variable as the spreadsheet software will allow. Strictly, the name is a formula that references the required cell or range and, in doing so, conceals the direct reference from the user. The name does not store the result; that is the province of the cells, which hold the calculated value as a property. The Name can, nevertheless, provide a meaningful and well-structured description of the cell's intended content. Normally, a Name will refer to a range as an absolute reference. In that instance, an inverse relationship will also exist whereby the Range object will have the Name as one of its properties.

There is, however, nothing that can be achieved using defined Names that could not also be achieved with direct referencing; it is a simple mechanical process to substitute each Name by its direct reference. All one loses by avoiding the use of Names is structure and meaning. Formulas will no longer be expressed in terms of the business rules they implement but will instead merely reference values by location.

Whereas direct cell referencing provides the means to address any of the 16 billion or so cells on the worksheet, the use of Names limits the address space to the previously declared ranges, in doing so reducing risk. Used in this manner, Names provide the functional equivalent of the VBA

```
Option Explicit
```

followed by

```
Dim myArray(1 To 100, 1 To 2) As Variant
```

which is generally considered good practice within the developer community.

In fact though, there is significant resistance to the use of defined names in the Excel community with a number of well-known publications in which the authors deprecate their use. When single-cell naming practices were studied by McKeever & McDaid, using names like "HMV2009Profits" it was found that they tend to have a detrimental effect on both the ease of creating a spreadsheet solution and upon the detection of errors, although the latter conclusion was originally drawn using inexperienced Excel users. In the opinion of the present author, the idea of having names in one-to-one correspondence with used cells renders them practically valueless; the number of names will be so large that they have little more significance than the cell references they replace and the simplest of tasks such as summing a contiguous range would become a nightmare.

Names come into their own in that they provide structure to the solution process by identifying large blocks of numbers that together represent features of the problem to be addressed at a more abstract conceptual level than may be achieved by referencing individual cells. The number of Names that need to be understood to develop a solution can be orders of magnitude fewer than the count of used cells, so should be far more memorable. The use of names also permits the more complex methods of defining ranges to be considered in order to capture the structure of the problem being addressed. For example, in the similar line item model shown in Section 2.2, the entire model timeline fell within a band of complete columns defined to be "*model*" that refers to

= $F:$X

The range representing a single period would be found by using MATCH to give an index value "*selectedPeriod*". In this case, the range "*inPeriod*" comprising the single column containing data for that period would be given by

= INDEX( *model*, 0, *selectedPeriod* )

If one were to have the revenue per period for a list of products identified by the range "*revenue*", the intersection

= (*revenue inPeriod*)

would be a fully dynamic range containing all the sales values for the chosen period and

= SUM(*revenue inPeriod*)

would be the total revenue for the period across all line-items. The same result might equally be achieved using direct cell referencing but it lacks the simplicity and readability of the named version.

A key aspect of the use of Names is that they **define** the Objects, simultaneously capturing meaning as well as defining their location and extent, whereas labels merely annotate with the intention of informing the reader as to the intended significance of adjacent data but they lack any direct linkage to it. The data is only identified though the observance of stated or assumed presentation standards which is a vastly weaker concept. A Name also has the advantage of being computer intelligible; it is part of the Names collection it can be accessed by computer program and the content can be read and processed programmatically, as will be shown in Section 4.

In the opinion of the author, Names are grossly underused but there is hope; at least one company (a firm of actuaries) has implemented an automated system that rejects any model that has even one cell address reference as part of its quality control process.

### 2.4 Array formulas

It is often claimed that array formulas are "powerful" and they certainly carry the stigma of being "advanced". To my mind that is an erroneous perception on both counts. The output of any array formula is defined in terms of the cell-by-cell calculations it represents. Provided one is willing to use helper cells to store intermediate results, it is always possible to replace an array formula by the cell-by-cell formulas that define it. Array formulas are not powerful, they only appear that way because they can hold intermediate results in memory, so making it seem that a final aggregation appears as if by magic. If one wishes to examine the intermediate calculation, the values generated may be output as an array formula anywhere on any worksheet without regard to relative positioning.

What array formulas really achieve, is an order of magnitude reduction in the number of independent formulas that are used to build a solution and, related to that, they provide a heavily constrained and restricted solution process. At first sight, it might appear that this loss of flexibility is a major handicap for anyone trying to build a solution from array formulas. In practice, it is usually the case that valid solutions conform to the constraints of array formulas whilst errors violate the constraints and require the use of single-cell formulas and relative references. By way of example, in Section 2.2 all 228 cells of the product revenue range are given by the single formula

{= *volume* * *product.price*}

Even more challenging, some line items are specified as having prices that escalate over time whilst others are constant. Exception handling is required both for the initialisation at the first period and to eliminate the escalation where it is not required. Whereas single-cell strategies allow the formula to be varied from cell to cell, the use of an array formula demands that the exception handling is built into the formula. Thus the '*isEscalated*?' flag is an array that applies item by item whilst '*initialise*?' applies to the first period only. The range '*←price*' is identical in size to '*product.price*' but displaced one cell to the left. The result is an accumulation that runs left to right across the entire range

```
{= IF( isEscalated?,
      IF( initialise?, price.initial, ←price) * ( 1 + price.escalationPerPeriod ),
      price.initial )}
```

## 2.5 Formatting and Graphics

My expectation of a model built upon Excel as a platform is that it should allow the client to explore a business scenario; it is not created in order that the user should be able to demonstrate their Excel skills; there may be no requirement even to expose the Excel development interface. In the World Cup predictor application, below, hyperlinked Group tabs replace the normal sheet tabs, the grid has no role in the location of cells and is not visible and the Excel ribbon is hidden.

The worksheet is designed to provide a graphic user interface and it is hoped that the user will have an intuitive understanding of its operation. This concept of 'affordance' was introduced by Norman, [1988] in the context of the design of everyday objects and applied to spreadsheets by Hellman [2005]. This design has achieved its objective if the user is able to move between Groups by clicking the appropriate tab and understands that they are to complete the scores either as a prediction or following the matches. The colour palette is based upon the national flag of the host nation Brazil but that is a purely an aesthetic consideration.

The user is not expected to interact with the application in the terms of an Excel workbook, nor examine the logic of the tiebreak algorithm; they simply see the result as the Group winner and runner-up names and their associated flags and see the teams transferred to the sheet representing the knockout stage of the competition.

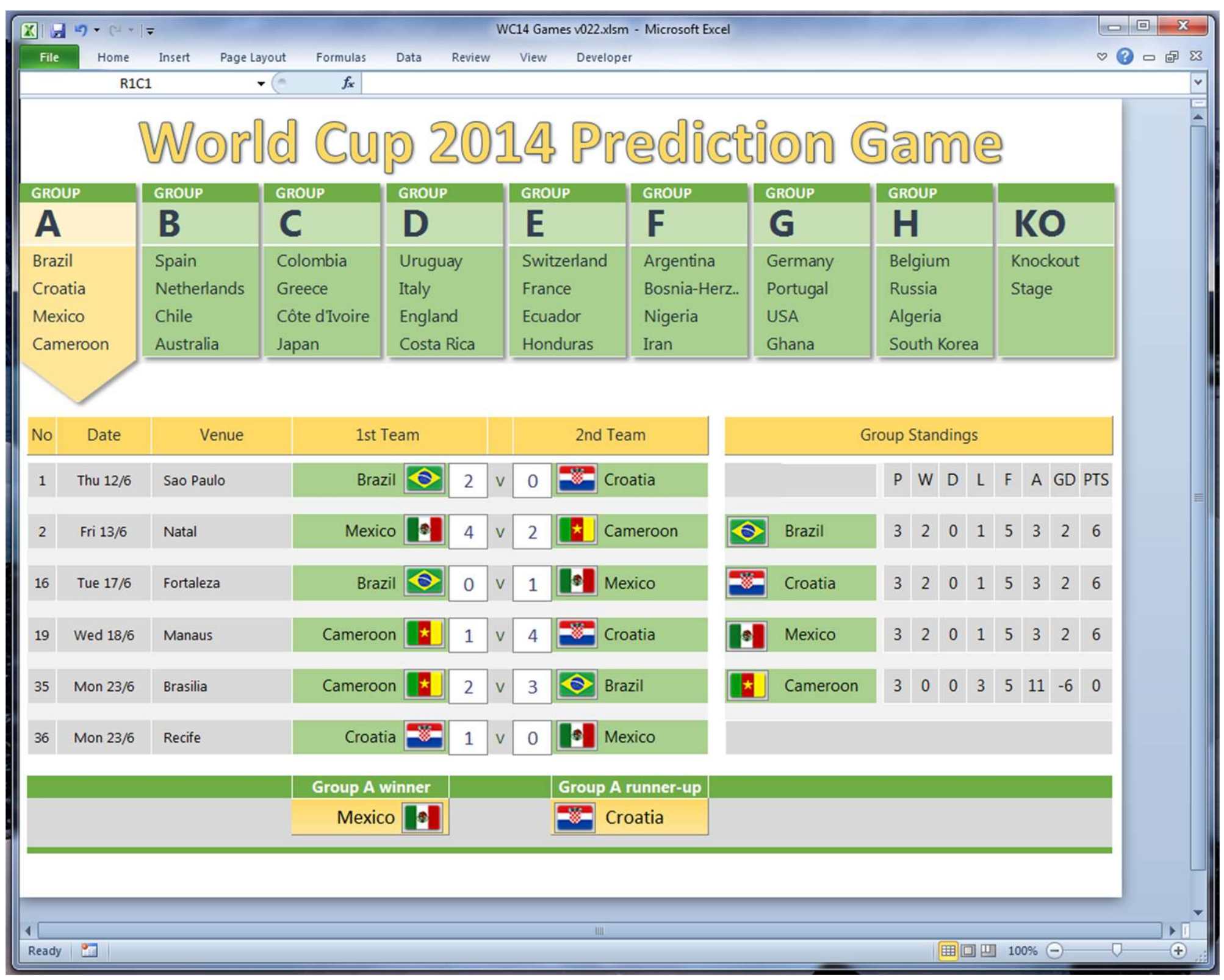

**Figure 2 Group stage sheet from the World Cup predictor workbook**

Microsoft offers two very different styles of default format. The most familiar is the grid of grey lines forming an array of white cells limited only by the worksheet boundaries. The more recent exemplar is the gaily-coloured horizontal stripes of the Table object with contrasting headings and filter dropdowns.

It may be argued that both represent an element of over-formatting; after all white space is not formatted as cells in any other media form and the positive formatting used to outline empty cells merely distracts attention from the significant content appearing elsewhere within the worksheet within data objects. Raffensperger, however, argues that blank cells are not the equivalent of graphic blank space because there is no guarantee that other cells do not reference it and the cell may contain hidden content. While it is clear that the developer can use formatting to conceal information in this manner, it is far more productive to consider how well-chosen formatting may best be used to convey information.

I personally favour using the Excel "normal style" to invoke a background 'colour wash' that eradicates all indication of cells. In that way, the developer is almost forced to format any cells they wish to include as part of the model. Most users find this very disconcerting at first because they are conditioned to think in terms of the grid system to identify individual cells and expect to see clear links between the content and the sheet headers. Once one dispenses with direct cell referencing, however, the ability to locate individual cells by coordinates is simply not required; the content is, instead, identified by Name and index within large, structured data Tables and Ranges.

Since the proposed development practice requires all referencing to be done through the use of named ranges, the cell-formatting could be applied automatically by scanning through the names collection using VBA utilities to identify data cells or formula ranges and formatting them using the appropriate style. I tend to favour a white background for input cells but might also introduce a white fill for output since it is these categories of information are of importance to the user and should appear with the greatest clarity. A font difference could be introduced to distinguish the two especially if they appear in close proximity. The intermediate formula cells are formatted with a grey fill to emphasise their secondary role as well as to discourage attempts to change these cells which are normally protected when the workbook is in use.

Equally, the Excel Tables may be regarded as over-formatted to the point of being lurid. Whilst the tables themselves stand out, the content is not as easy to read against a coloured background as it is against white. The direction of the stripes may be relevant as a device for identifying the properties of a record within a list but is less appropriate for a horizontal timeline or a matrix. Similarly, the concept of a row filter is not universally applicable.

## 3 NAMED FORMULA AS AN AID TO ENCAPSULATION AND REUSE

Names provide clear-cut interfaces between one functional unit of spreadsheet coding and the various functions that reference the data. There is, however, nothing to stop the developer from completely reworking the calculation methodology, provided the meaning is maintained and the datatype is unaltered.

For example, some taxes like sales tax in the US or VAT in the EU vary by jurisdiction. It is quite possible that a spreadsheet solution originally intended to work within a single jurisdiction might eventually need to be generalised to cope with varying values. An absolute reference to a single cell

| Name | Refers to (absolute range) |
|---|---|
| *taxRate* | = $J$16 |

could be replaced by some form of LOOKUP (one of a number of possible implementations) such as

| Name | Refers to (formula) |
|---|---|

| | | = LOOKUP( *state*, *state.codes*, *state.taxRate* ) |
|---|---|---|
| *taxRate* | or | = INDEX( *state.taxRate*, MATCH( *state*, *state.codes*, 0 ) ) |

to create an array of values, one for each line-item. Any array formula that references *taxRate* will adapt automatically. Interestingly, a named formula may also provide a simple method of regressing the workbook functionality to its legacy mode if that were still required. Just slightly more complicated is

| Name | Refers to (formula) |
|---|---|
| *taxRate* | = IF( *isLegacyCalculation?*,<br>*default.taxRate*,<br>LOOKUP( *state*, *state.codes*, *state.taxRate* )<br>) |

The named formula provides a single point of access to correct or enhance the functionality of the spreadsheet solution by encapsulation of the means of its calculation. Had every tax calculation formula within the workbook made direct reference to the cell containing the original tax rate, the process of updating the workbook would require the addition of a new helper range and would have incurred significantly greater risk.

A more drastic form of modular reuse is made possible by the embedding a copy of a worksheet taken from a template workbook to provide new functionality, as illustrated in the following section. The formulas embedded within the imported worksheet may be viewed in the same light as a call to a library in a procedural code or an additional method within OO programs.

### 3.1 Merging two sorted lists using an imported Excel module

This example illustrates the use of Names to allow the import of a 'black box' module from a separate workbook. The main window of Figure 3 shows the master sheet whilst the module is superposed on it. The key element of the strategy is that the module contains only Names scoped to the worksheet that is to be imported. The private properties '*private.list.A*' and '*private.list.B*' are initially linked to dummy data to allow the module to be unit tested independently. By redefining these Names to refer to the public variables held in the master document links are made that achieve the same objective as the Lets and Sets of the Object-Oriented paradigm. Similarly the equivalent of a Get to return the result is achieved by setting the workbook name '*merged.list*' to refer to the module name '= *mergeRoutine!value*' essentially mimicking the 'value' property of a Class.

The purpose in presenting this example is to demonstrate the way in which well-chosen Names can be used to encapsulate both methods and properties (strictly just range values and formulae). The objective of allowing the functionality of the master workbook to be modified and further developed is far more readily achievable if the existing solution may be reused with little change. The meaning and values of the Names can be changed without requiring changes to be made to the formulas that reference them.

Some degree of caution in handling Names is required though. In particular, it should be recognised that deleting a worksheet to replace it with a fresh version does not remove the Names defined as being local to the sheet [Grossman & Burd, 2015]. This can create difficulty, so it is recommended that the local Names are all deleted before one attempts to delete the worksheet.

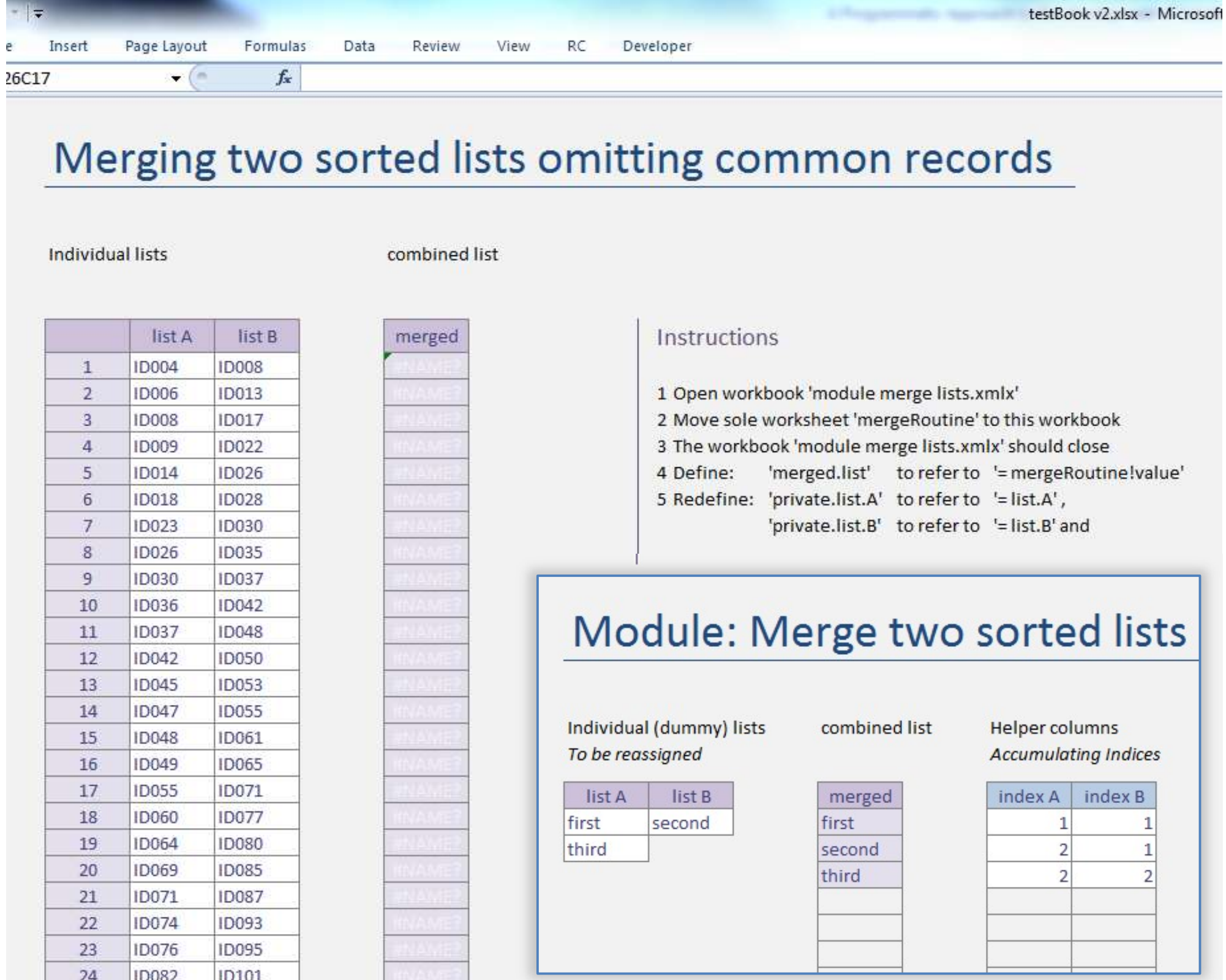

**Figure 3 The use of Names to provide Public and Private versions of parameters**

What is vital is that every defined Name should be local to the imported worksheet, otherwise undesirable and confusing crosslinks between workbooks will result. Once the new workbook is in place, the local Names used for data input would be reassigned to refer to the appropriate global Names from the host workbook (previously they may have been used to refer to dummy data held within the worksheet so that the functionality could at least be demonstrated). If the calculation simply defines a sequence of array formulas performed in memory there is no problem with needing to access the worksheet holding the new module (it could even be imported as a hidden worksheet) but it is more likely that some scratch workspace would be required in the form of a worksheet range. Whereas named formula will be sized entirely by the referenced input data arrays, any formula held within a range can only expand until it reaches the bounds imposed by the size of range. The presence of cells containing #N/A at the end of the helper ranges should not be a problem either in terms of aesthetics or performance but some care is needed to ensure the developer is warned if the size allocated is insufficient.

Just as the input links are made by allowing the local name to refer to the global name, there must be a suitable mechanism for returning the data objects that result from the calculation. The recommended strategy is simply to document the local names used for the output. Assuming the sheet name is readily associated with the new functionality, the reference to the output using the full name including the sheet name should be acceptable. Other names which should not be touched could be hidden for their protection. Unfortunately Excel does not provide any means of protecting defined names so it may be useful to rebuild the names from documentation if there is any suspicion that any defined names may have been corrupted.

## 4 AUDITING AND AUTOMATION

### 4.1 Example: Standard loan repayment model

This example is based upon the FAST modelling guide [Sternberg] that describes how their methodology may be applied to a loan calculator with a choice of debt repayment profiles and includes the possibility of a period of grace in which the borrower is obliged neither to pay debt interest nor principal.

Whilst the present model follows the commentary very closely, basing the name selection upon the text description, it departs from the FAST standard in several respects.

Loan Calculator Variable model 4A.xlsm - Microsoft Excel

File Home Insert Page Layout Formulas Data Review View RC Developer Add-Ins

R17C66 {= IF( initialise.loan?, loan.amount, ←debt.balance + interest.expense - debt.service )}

**Loan calculations**

*Time Line*

| | | | | | | | | | | | | | |
|---|---|---|---|---|---|---|---|---|---|---|---|---|---|
| start of period | date | 01 Nov 19 | 01 Dec 19 | 01 Jan 20 | 01 Feb 20 | 01 Mar 20 | 01 Apr 20 | 01 May 20 | 01 Jun 20 | 01 Jul 20 | 01 Aug 20 | 01 Sep 20 | 01 Oct 20 |
| end of period | date | 30 Nov 19 | 31 Dec 19 | 31 Jan 20 | 29 Feb 20 | 31 Mar 20 | 30 Apr 20 | 31 May 20 | 30 Jun 20 | 31 Jul 20 | 31 Aug 20 | 30 Sep 20 | 31 Oct 20 |
| **Calculation** | | | | | | | | | | | | | |
| *Event and Status Flags* | | | | | | | | | | | | | |
| Key: Initialise loan (as start of next period) | ⇧ | REPAYMENT | REPAYMENT | ⇩ | | | | | | | | | |
| Grace period ends (ie first repayment period) | ⇨ | REPAYMENT | REPAYMENT | ⇩ | | | | | | | | | |
| Final repayment period (should show zero) | ⇩ | REPAYMENT | REPAYMENT | REPAYMENT | REPAYMENT | REPAYMENT | REPAYMENT | REPAYMENT | ⇩ | | | | |
| Grace and repayment periods | GRACE | REPAYMENT | REPAYMENT | REPAYMENT | REPAYMENT | REPAYMENT | REPAYMENT | REPAYMENT | ⇩ | | | | |
| **Overall Balance** | | | | | | | | | | | | | |
| *Debt balance* | | | | | | | | | | | | | |
| Debt balance Fixed Principal No Grace | USD | 33,333 | 16,667 | - | - | - | - | - | - | - | - | - | - |
| Debt balance Fixed Payment No Grace | USD | 48,719 | 24,601 | - 0 | - 0 | - 0 | - 0 | - 0 | - 0 | - 0 | - 0 | - 0 | - 0 |
| Debt balance Fixed Principal 6 Months Grace | USD | 122,618 | 105,101 | 87,584 | 70,067 | 52,551 | 35,034 | 17,517 | - | - | - | - | - |
| Debt balance Fixed Payment 6 Months Grace | USD | 174,356 | 150,903 | 126,981 | 102,580 | 77,692 | 52,306 | 26,412 | - 0 | - 0 | - 0 | - 0 | - 0 |
| **Cost elements** | | | | | | | | | | | | | |
| *Interest paid* | | | | | | | | | | | | | |
| Interest paid Fixed Principal No Grace | USD | 1,000 | 667 | 333 | - | - | - | - | - | - | - | - | - |
| Interest paid Fixed Payment No Grace | USD | 1,447 | 974 | 492 | - 0 | - 0 | - 0 | - 0 | - 0 | - 0 | - 0 | - 0 | - 0 |
| Interest paid Fixed Principal 6 Months Grace | USD | 2,803 | 2,452 | 2,102 | 1,752 | 1,401 | 1,051 | 701 | 350 | - | - | - | - |
| Interest paid Fixed Payment 6 Months Grace | USD | 3,947 | 3,487 | 3,018 | 2,540 | 2,052 | 1,554 | 1,046 | 528 | - 0 | - 0 | - 0 | - 0 |
| *Principal repaid* | | | | | | | | | | | | | |
| Principal repaid Fixed Principal No Grace | USD | 16,667 | 16,667 | 16,667 | - | - | - | - | - | - | - | - | - |
| Principal repaid Fixed Payment No Grace | USD | 23,645 | 24,118 | 24,601 | - | - | - | - | - | - | - | - | - |
| Principal repaid Fixed Principal 6 Months Grace | USD | 17,517 | 17,517 | 17,517 | 17,517 | 17,517 | 17,517 | 17,517 | 17,517 | - | - | - | - |
| Principal repaid Fixed Payment 6 Months Grace | USD | 22,993 | 23,453 | 23,922 | 24,400 | 24,888 | 25,386 | 25,894 | 26,412 | - | - | - | - |
| *Debt service* | | | | | | | | | | | | | |
| Debt service Fixed Principal No Grace | USD | 17,667 | 17,333 | 17,000 | - | - | - | - | - | - | - | - | - |
| Debt service Fixed Payment No Grace | USD | 25,093 | 25,093 | 25,093 | - | - | - | - | - | - | - | - | - |
| Debt service Fixed Principal 6 Months Grace | USD | 20,320 | 19,969 | 19,619 | 19,269 | 18,918 | 18,568 | 18,218 | 17,867 | - | - | - | - |
| Debt service Fixed Payment 6 Months Grace | USD | 26,940 | 26,940 | 26,940 | 26,940 | 26,940 | 26,940 | 26,940 | 26,940 | - | - | - | - |
| *Insurance cost* | | | | | | | | | | | | | |
| Insurance cost Fixed Principal No Grace | USD | 25 | 17 | 8 | - | - | - | - | - | - | - | - | - |
| Insurance cost Fixed Payment No Grace | USD | 36 | 24 | 12 | - 0 | - 0 | - 0 | - 0 | - 0 | - 0 | - 0 | - 0 | - 0 |
| Insurance cost Fixed Principal 6 Months Grace | USD | 70 | 61 | 53 | 44 | 35 | 26 | 18 | 9 | - | - | - | - |
| Insurance cost Fixed Payment 6 Months Grace | USD | 99 | 87 | 75 | 63 | 51 | 39 | 26 | 13 | - 0 | - 0 | - 0 | - 0 |

Inputs SeriesCalcs Outputs

Ready

**Figure 4 Final stages of four alternative debt repayment profiles**

As already described, the output of this workbook is intended to show the salient features of the loan repayment but with no allusion to the Excel grid or cell references. The user is simply not expected to inspect the technicalities of the calculation.

In fact the formula bar shows the Debt Balance as being calculated using the formula

```
{= IF( initialise.loan?,
       loan.amount,
       ←debt.balance + interest.expense - debt.service )}
```

which may be reasonably meaningful to someone who has no knowledge of Excel. It may be more of a challenge for a reviewer that attempts to use conventional auditing techniques. Even the use of the symbols: "←" (back arrow) to indicate the previous period; "." (full stop) to distinguish some property of the root name; and "?" to indicate a Boolean flag will initially appear alien.

### 4.2 Validating formulas without direct cell referencing

One of the criticisms made of the use of Names is that they insert an additional layer of indirection that makes it harder for the reviewers to navigate to the source of referenced data. That is undoubtedly true but the question is, "Does that matter?" In developing, verifying or auditing traditional workbooks we are conditioned to accept the need to branch about the workbook in the hope that the intent of the formula we are evaluating will be clearer once we see the cells that it references. The trouble is that content of referenced cell may be as obscure as the original in providing meaning. In a well-ordered workbook it is likely that there will be some form of annotation or table heading that may give a clue to the meaning of the referenced cell but that is far from universally true and it is very easy to destroy the original order of lists in workbooks based upon direct single-cell references or to separate items from the intended annotation.

By comparison, validating a workbook based upon defined names can be a very unfamiliar process. There are plusses though; the number of Names should be orders of magnitude fewer than the number of used cells. Also, provided the cells containing a worksheet formula are always defined as a named range, it is possible to create a complete description of the workbook formulas using Name Manager or other, improved, versions of the tools. Used with care, the names pasted to the worksheet should provide a reasonably good view of the flow of calculation through a sequence of readable and semantically meaningful formulae.

Better still is to use a VBA utility to parse the formulas and use the resulting predecessor graph to order the output for presentation as a navigable diagram.

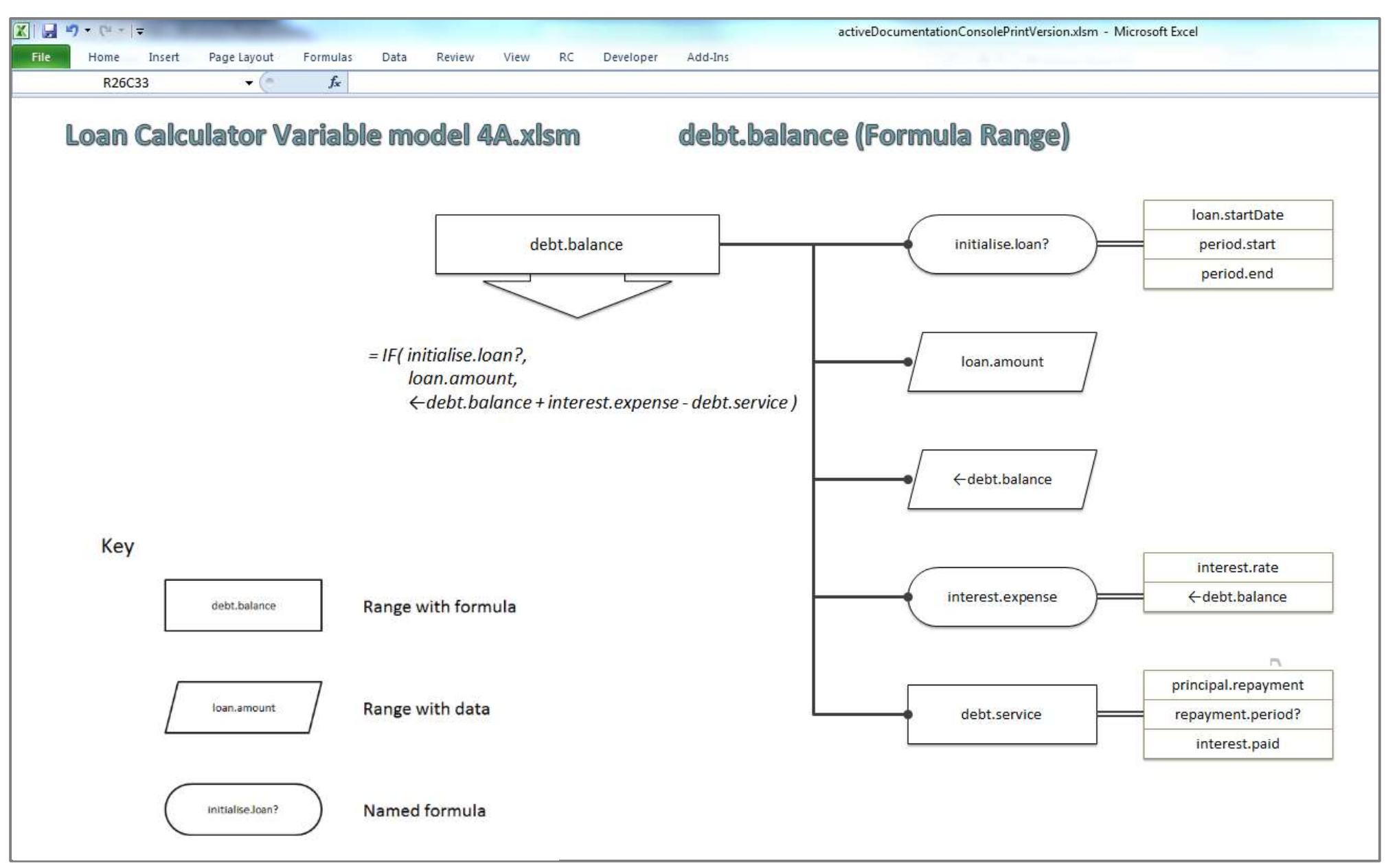

**Figure 5 Graph showing the references and usage of the 'debt.balance' formula.**

In Figure 5, the diagram is set with its focus upon the debt balance and shows the formula contained within the range. The connectors provide links to the ranges referenced within the formula. It may appear a little surprising that the 'debt.balance' should appear with no dependents (*i.e*. nothing is shown to the left as referencing it). That is simply because the balance references are all, in fact, to the previous period's balance '←debt.balance', as shown in Figure 6.

The other element of evaluating the spreadsheet solution is to check that the Named Ranges actually contain the data that one is led to expect from the name; the spectre of a name being misapplied is often raised as an issue, though I would argue the problem is the other way round. The Name defines the data structure; it may be incorrectly dimensioned but it cannot, by definition, be incorrectly placed. The question should be, "Given the Named Range, is the data appropriately entered?" This documentation tool addresses the problem by providing the 'down arrow' control, which takes one to the target workbook with the relevant range selected on the worksheet.

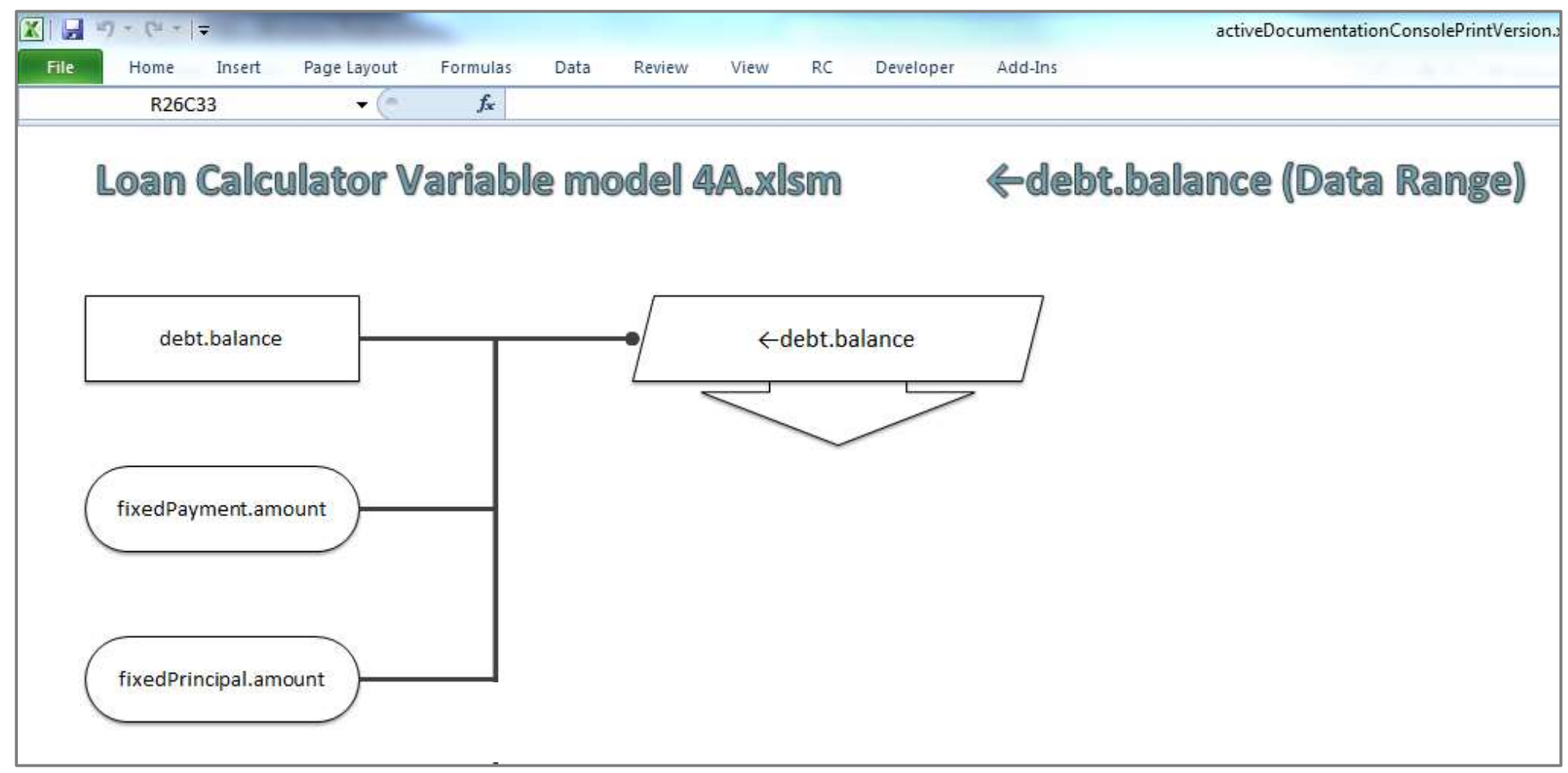

**Figure 6 Graph showing the usage of the offset range ←*debt.balance*.**

## 5 CONCLUSIONS

The point of this paper is not to suggest that developers should change the habits of a lifetime and adopt the strategy as outlined. It is more the intention to raise the awareness that the adoption of less common approaches within Excel can lead to a coherent solution strategy in which the problem is solved by a sequence of formulas resembling the steps of a programmed language. Individually the techniques can deliver benefit when building models and the developer ought to be aware of the possibilities even though they may choose to stay with default techniques that will be more familiar to their clients.

The author has developed spreadsheet solutions with array formula and named ranges for a number of years now but the experience within the community of the risks of such an approach is miniscule and the scarcity of Excel specialists capable of maintaining such workbooks is itself a risk that must be mitigated. It is certain that the end user is likely to have far less opportunity to 'meddle' but, there, improved workbook integrity must be balanced against possible client dissatisfaction.